\documentclass[letterpaper]{article}
\usepackage{aaai2026} 
\usepackage{times}
\usepackage{helvet}
\usepackage{courier}
\usepackage[hyphens]{url}
\usepackage{graphicx}
\usepackage{natbib}
\usepackage{caption}
\usepackage{algorithm}
\usepackage{algorithmic}
\usepackage{amsmath}
\usepackage{amsfonts}
\usepackage{amssymb}

\usepackage{newfloat}
\usepackage{listings}
\usepackage{ragged2e}
\usepackage{booktabs}
\usepackage{pifont}
\usepackage{xcolor}

\DeclareCaptionStyle{ruled}{labelfont=normalfont,labelsep=colon,strut=off}
\floatstyle{ruled}
\newfloat{listing}{tb}{lst}{}
\floatname{listing}{Listing}

\title{HemePLM--Diffuse: A Scalable Generative Framework for Protein--Ligand Dynamics in Large Biomolecular System}
\author{
    Rakesh Thakur\textsuperscript{\rm 1},
    Riya Gupta\textsuperscript{\rm 1}
}
\affiliations{
    \textsuperscript{\rm 1}Amity Centre for Artificial Intelligence, Amity University, Noida\\
    rakeshthakur35016@gmail.com, riyagupta2231@gmail.com
}

\begin{document}
\nocopyright
\maketitle

\begin{abstract}
Comprehending the long-timescale dynamics of protein--ligand complexes is very important for drug discovery and structural biology, but it continues to be computationally challenging for large biomolecular systems. We introduce HemePLM-Diffuse, an innovative generative transformer model that is designed for accurate simulation of protein--ligand trajectories, inpaints the missing ligand fragments, and sample transition paths in systems with more than 10,000 atoms. HemePLM-Diffuse has features of SE(3)-Invariant tokenization approach for proteins and ligands, that utilizes time-aware cross-attentional diffusion to effectively capture atomic motion. We also demonstrate its capabilities using the 3CQV\_HEME system, showing enhanced accuracy and scalability compared to leading models such as TorchMD-Net, MDGEN, and Uni-Mol.
\end{abstract}

\section{Introduction}

Simulating the long-timescale behaviour of protein--ligand systems is a very fundamental task in computational biology, essential for the better understanding of binding mechanisms, enzyme catalysis, and allosteric regulation. These simulations offer very high-resolution insights into the dynamic processes that drive biomolecular functions and are extensively utilized in drug discovery and structural prediction workflows. The Traditional Molecular Dynamics (MD) simulations, that are conducted with software such as GROMACS or AMBER, rely on the numerical integration of Newton’s equation of motion. Although these methods are very accurate, they can also become computationally demanding when applied to very large systems or when capturing rare events, such as ligand dissociation or conformational transitions, which can occur over microsecond to millisecond timescales \cite{KarplusMcC2002}.

In the very recent years, deep learning-based approaches have emerged as promising alternatives to the Molecular Dynamics (MD). Models such as TorchMD-Net \cite{Tholke2021} and SchNet \cite{Schutt2017} focus on learning potential energy surfaces or force fields directly from atomic structures. More recently, MDGEN \cite{Chen2023MDGEN} has introduced a diffusion-based generative model to generate realistic protein dynamics without the need for expensive integration. In addition to that, pretraining frameworks like Uni-Mol \cite{Zhou2023UniMol} offers SE(3)-equivalent embeddings of static molecular structures, although they do not simulate their temporal evolution.

Despite these advancements, the current models face a very significant limitation: they primarily address small, isolated systems and struggle to scale to large, all-atom protein--ligand complexes, particularly those with flexible binding sites and multiple metastable states. For example, TorchMD-Net \cite{Tholke2021} mainly supports small proteins without ligands, while The MDGEN \cite{Chen2023MDGEN} focuses only on coarse-grained backbone dynamics (e.g., C$\alpha$ Traces). Although Uni-Mol \cite{Zhou2023UniMol} excels in molecular property prediction, it lacks a generative temporal module and is not intended for trajectory simulation.

To address the limitations of the existing models, we are here to present HemePLM-Diffuse, a scalable and versatile generative model designed for simulating the long-timescale dynamics of large protein--ligand systems. This model has features of SE(3)-invariant tokenization for both protein and ligand atoms, ensuring adherence to the geometry of a 3D molecular space. It employs a time-aware cross-attention transformer architecture that effectively models all-atom interactions over extensive spatiotemporal scales. In addition to that, HemePLM-Diffuse incorporates a diffusion-based trajectory inpainting and transitions path sampling. Our evaluation of this experiment on a real-world system (3CQV\_HEME) with over 10,000 atoms demonstrates capabilities far beyond those of the current models. This work marks a significant advancement toward a generalizable, end-to-end protein--ligand dynamics simulator that is both fast and physically grounded, paving the path for new applications in structure-based drug design, protein engineering and molecular generative modeling at unprecedented scales.

\section{Related Work}

To understand and model the molecular dynamics, particularly in large protein--ligand systems, sits at the intersections of physics-based simulation, equivariant modeling, and generative deep learning, We categorize this relevant work into five broad areas.

Equivariant models for Molecular Properties and Dynamics: Initial deep learning efforts in the molecular modeling puts the light on the equivariant neural force fields such as TorchMD-Net \cite{Tholke2021}, SchNet \cite{Schutt2017}, and PaiNN \cite{Schutt2021PaiNN}. These models have successfully learned the potential energy surfaces and predicted forces while respecting the SE(3)-equivariance.

However, their dependence on the explicit numerical integration for the dynamics that led to the computational bottlenecks and error accumulation, that is restricting them to small systems that have less than 1000 atoms, without explicit ligand interactions or long-range transitions.

Simultaneously, the 3D molecular pre-training models such as Uni-Mol \cite{Zhou2023UniMol}, Molformer \cite{Ross2022}, and GeoMol \cite{Ganea2021} demonstrated the power of the transformer backbones and GNNs for learning the molecular representation and the properties for the static 3D structures. The extensions like EquiBind \cite{Stark2022} and the DiffDock \cite{Corso2022} have enabled static docking predictions. While the powerful for molecular understanding and property prediction, these models fundamentally lacked the ability to simulate the dynamic temporal evolution or scale to full protien-ligand complexes.

The emergence of the Generative approaches for Dynamics: In order to overcome the limitations of integration-based methods, generative molecular dynamics frameworks began to emerge, aiming to directly sample plausible tracjectories. MDGEN \cite{Chen2023MDGEN}, for instance, pioneered the use of diffusion models for protein trajectories, and EnDiff \cite{Liu2024} introduces energy-guided diffusion for conformational landscapes. These methods offer faster sampling by bypassing expensive force calculations.Despite these advancements in the dynamic modeling, the crucial discrepancy remained: importantly focuses on coarse-gained features, neglected atomistic accuracy, or were confined to small molecules or isolated peptides in simplified conditions lacking solvent and complex biological interactions \cite{Tholke2021, Chen2023MDGEN, Zhou2023UniMol}.

HemePLM-Diffuse: To bridge the discrepancy with SE(3)-Invariance, our work is built on the fundamental and foundational principles of this invariant modeling which preserves the geometric relationships in the 3D data.

Most importantly, we adapt this to a time-aware generative transformer capable of modeling realistic dynamics for entire protein--ligand systems. To our knowledge, this is the first model to transcend the limitations of existing approaches by scaling to full all-atom protien-ligand systems with over 10,000 atoms. This comprehensive approach marks a significant advancement towards generalized, end-to-end protein--ligand dynamics simulations.

\section{Methods}

\subsection{Dataset Construction and System Selection}

\begin{figure}[htbp]     
\centering
\includegraphics[width=1.0\columnwidth]{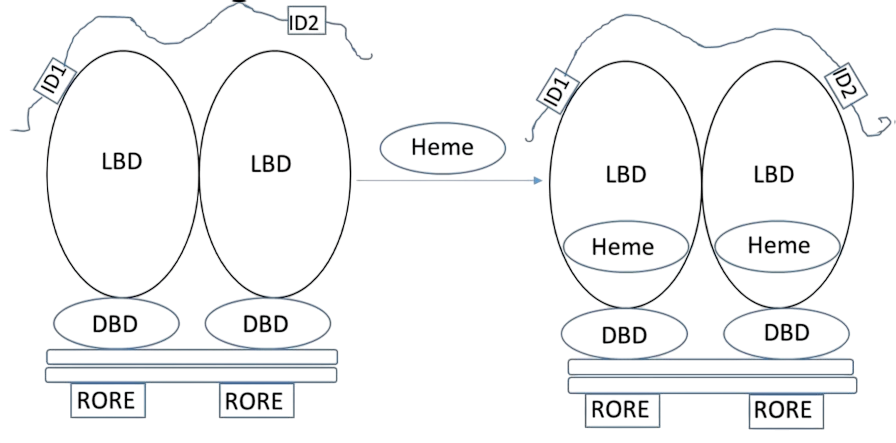}
\caption{The Rev-erb$\beta$ nuclear receptor binding mechanism. The diagram shows the LBD (Ligand Binding Domain), DBD (DNA Binding Domain), and RORE (Retinoic acid-related orphan receptor response element) regions of the receptor, and how the Heme ligand binds to the LBD region, inducing a conformational change.}
\label{fig:binding_mechanism}
\end{figure}

We curate a dataset of a large protein--ligand complex from the Protein Data Bank (PDB), which focuses on systems with at least 10,000 atoms, bound ligands of variable topology, and high-resolution structures ($\le$2.5 \AA). For our experiments, we use PDB ID: 3CQV, which represents the nuclear receptor Rev-erb$\beta$ that binds to the HEME ligand. This complex is solvated and equilibrated using GROMACS \cite{VanDerSpoel2005} with the Amber99SB-ILDN force field, and atomic coordinates are extracted for downstream modeling.

\subsection{SE3-Invariant Tokenization}
Inspired by SE(3)-Transformers \cite{Fuchs2020} and Tensor Field Networks \cite{Thomas2018}, we tokenize proteins and ligands using rotation- and translation-invariant features. The protein tokens consist of the 3D coordinates of C$\alpha$ atoms, and the backbone torsion angles, and auxiliary features such as B-factor, SASA, and binding site scores from AlphaFold \cite{Jumper2021}. The ligand tokens include a fragment-based representation via RDKit \cite{Landrum2020}, relative 3D positions to protein residues, atom-wise descriptors, internal dihedrals, and top-k residue-ligand distances. This results in two tensors: $P \in \mathbb{R}^{L_p \times d_p}$ for proteins and $L \in \mathbb{R}^{L_l \times d_l}$ for ligands.

\subsection{Generative Trajectory Transformer}

The model takes as input noised tokens, denoted as $(P, L)$, at a specific time step $t$. Additionally, it can incorporate optional conditions such as the start and end conformations, partial ligand information, and a contact map.

\begin{figure*}[htbp]
\centering
\includegraphics[width=0.62\textwidth]{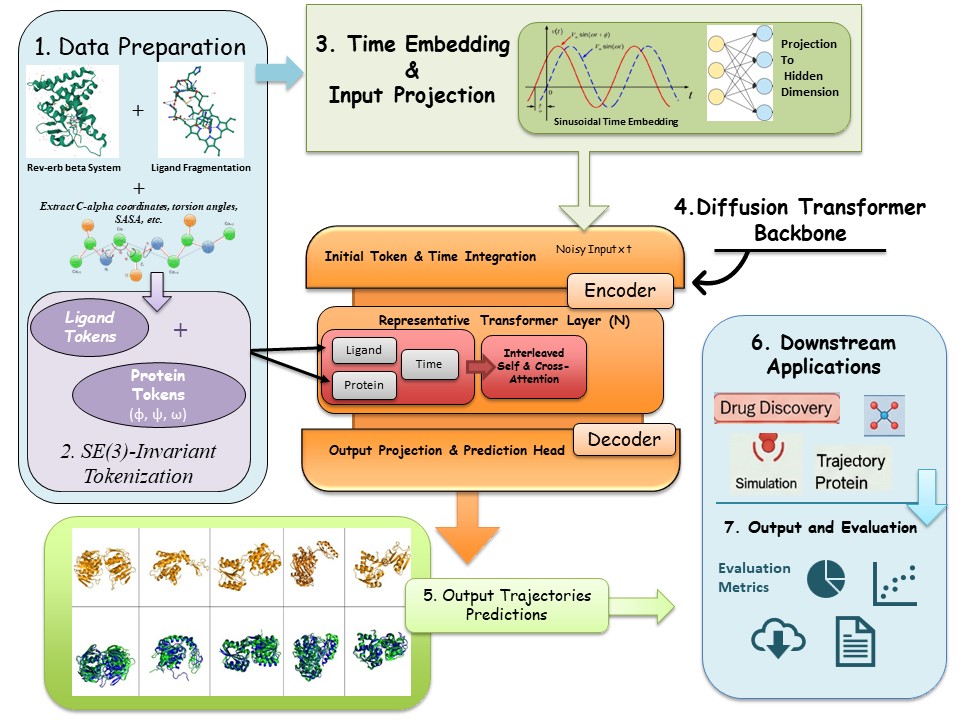}
\caption{{Architecture of HemePLM-Diffuse.} An overview of the HemePLM-Diffuse model. 1. Data Preparation involves processing the Rev-erb Beta system and fragmenting the ligand. 2. SE(3)-Invariant Tokenization generates Ligand Tokens and Protein Tokens ($\mathbf{\Phi}$, $\mathbf{\Psi}$, $\mathbf{\Omega}$). 3. Time Embedding \& Input Projection incorporates sinusoidal time embedding and projects input to a hidden dimension. 4. Diffusion Transformer Backbone uses an Encoder with Initial Token \& Time Integration, Representative Transformer Layers (N) with Interleaved Self \& Cross-Attention, and a Decoder for Output Projection \& Prediction Head. 5. Output Trajectories Predictions show the generated dynamic states. 6. Downstream Applications include Drug Discovery, Trajectory Simulation, and Protein analysis. 7. Output and Evaluation utilize various Evaluation Metrics.}
\label{fig:architecture}
\end{figure*}

The architecture employs an alternating mechanism of intra-molecule attention, ligand--protein cross-attention, and SE(3)-invariant multi-layer perceptrons (MLPs). This design enhances scalability and flexibility.

In terms of diffusion modeling, we utilize a stochastic velocity-based denoising approach, also known as flow matching. This is represented mathematically as shown in
\begin{equation}
x_t = x_o + \epsilon_t \label{eq:1}
\end{equation}
where $\epsilon_t \sim \mathcal{N}(0, \sigma^2)$. Our objective is to minimize the loss function $\mathcal{L}_{\text{traj}}$, defined as shown in
\begin{equation}
\mathcal{L}_{\text{traj}} = \mathbb{E}_{t} \left[ \left\| f_\theta(x_t, t) - \nabla_{x_t} \right\| \right] \label{eq:2}
\end{equation}
where $f_\theta$ predicts trajectory velocities over the time steps.

\subsection{Training Losses}
Our model optimizes a multi-objective loss function:

Trajectory MSE loss:
\begin{equation}
    \mathcal{L}_{\text{traj}} = \| \hat{P} - P \|^2 + \| \hat{L} - L \|^2 \label{eq:3}
\end{equation}

Free energy surface loss (via KL divergence between Markov state models):
\begin{equation}
    \mathcal{L}_{\text{FES}} = KL(P_{\text{model}} \| P_{\text{MD}}) \label{eq:4}
\end{equation}

Ligand inpainting loss (cross-entropy on masked fragments):
\begin{equation}
    \mathcal{L}_{\text{inpaint}} = -\sum_{i \in \text{masked}} \log p(L_i | P, L_{-i}) \label{eq:5}
\end{equation}

Transition path sampling loss:
\begin{equation}
    \mathcal{L}_{\text{TPS}} = \| x_t - \gamma(x_o, x_T, t) \|^2 \label{eq:6}
\end{equation}
Where $\gamma(\cdot)$ is a learnable interpolant conditioned on endpoints.

\subsection{Downstream Tasks Enabled}
Our model allows several downstream tasks that enhance the utility in molecular dynamics simulations. One of the primary capabilities is trajectory generation, which allows for the transition from an initial frame to a complete dynamic rollout of molecular movements. Additionally, the model supports trajectory upsampling, enabling the interpolation of sparse molecular dynamics sequences to create more detailed representations.

The other significant feature is ligand inpainting, which involves recovering missing atoms or fragments within a molecular structure. The model can also perform transition path sampling, simulating conformational switches, such as the transformation from an unbound to a bound state. Finally, it facilitates ligand optimization by suggesting dynamic edits that promote stable binding interactions. These functionalities together enhance the model's applicability in various molecular dynamics scenarios.

\section{Experiments}

\begin{figure*}[htbp]
    \centering
    \begin{minipage}[c]{0.45\textwidth}
        \centering
        \includegraphics[width=\textwidth]{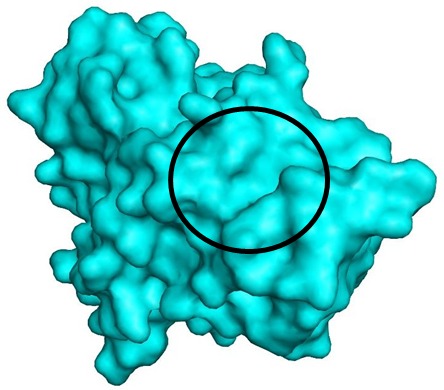}
        \caption{The 3CQV protein structure with the ligand binding pocket highlighted in a black circle.}
        \label{fig:protein_pocket}
    \end{minipage}
    \hspace{0.05\textwidth}
    \begin{minipage}[c]{0.45\textwidth}
        \centering
        \includegraphics[width=\textwidth]{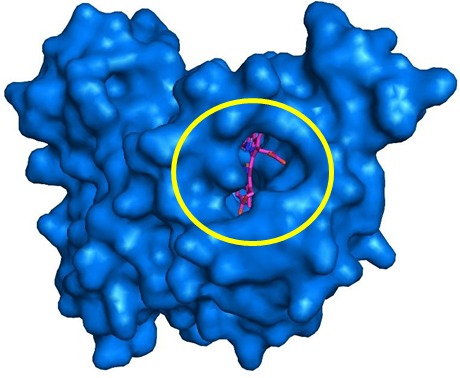}
        \caption{The 3CQV protein structure with the HEME ligand docked in the binding pocket, highlighted by a yellow circle.}
        \label{fig:protein_ligand}
    \end{minipage}
\end{figure*}

We evaluate HemePLM-Diffuse on the 3CQV system: the crystal structure of the Rev-erb$\beta$ nuclear receptor which is bound to a heme (HEME) ligand \cite{Yin2007}. This structure is selected due to its biological importance in circadian rhythm regulation and its challenge as a large system with ligand coupling. The complex includes over 1,700 atoms and represents a practical large-scale protein--ligand simulation benchmark.

\subsection{Setup}

For our experiments, we use the 3CQV+HEME complex, which has over 1,700 atoms. All experiments are conducted on an NVIDIA A100 GPU. One of the primary tasks involves ligand inpainting, where we randomly mask 10 atoms from the HEME structure and subsequently compute the root mean square deviation (RMSD) to evaluate the accuracy of the recovery. Additionally, we perform trajectory prediction by generating complete trajectories and calculating the RMSD in comparison to the molecular dynamics (MD) ground truth. Another important task is transition path sampling (TPS), where we sample the transition from bound to unbound states and assess the overlap between these states using Markov state models (MSM).

\subsection{Baseline Comparison Table}
\begin{table*}[htbp]
\footnotesize 
\begin{tabular}{p{1.6cm}|p{1.3cm}p{1.4cm}p{1.1cm}p{1.3cm}p{1.2cm}p{1.4cm}p{1.6cm}p{1.4cm}p{1cm}} 
\toprule
\textbf{Model} & \textbf{Type} & \textbf{Dynamics\newline Support} & \textbf{Ligand\newline Support} & \textbf{Scalability\newline ($>$1k atoms)} & \textbf{All-atom\newline Protein--Ligand} & \textbf{Ligand\newline Inpainting} & \textbf{Trajectory\newline Upsampling} & \textbf{Transition\newline Path Samp.} & \textbf{SE(3)-Equiv.\newline / Inv.} \\
\midrule
TorchMD-Net \newline (\cite{Tholke2021}) & SE(3)-equiv. MLFF & Num. Integration & \textcolor{red}{\ding{55}} & \textcolor{red}{\ding{55}} & \textcolor{red}{\ding{55}} & \textcolor{red}{\ding{55}} & \textcolor{red}{\ding{55}} & \textcolor{red}{\ding{55}} & \textcolor{green}{\ding{51}} \\
\midrule
MDGEN \newline (\cite{Chen2023MDGEN}) & Diffusion Gen. & Backbone only & \textcolor{red}{\ding{55}} & \textcolor{red}{\ding{55}} & \textcolor{red}{\ding{55}} & \textcolor{green}{\ding{51}} & \textcolor{green}{\ding{51}} & \textcolor{green}{\ding{51}} & \textcolor{red}{\ding{55}} \\
\midrule
Uni-Mol \newline (\cite{Zhou2023UniMol}) & 3D Graph Pretrain. & Static mol. & \textcolor{green}{\ding{51}} & \textcolor{red}{\ding{55}} & \textcolor{red}{\ding{55}} & \textcolor{green}{\ding{51}} & \textcolor{red}{\ding{55}} & \textcolor{red}{\ding{55}} & \textcolor{green}{\ding{51}} \\
\midrule
HemePLM-Diffuse \newline (Ours) & E(3)-inv. Diffusion & Full traj. \newline (XYZ + Torsion) & \textcolor{green}{\ding{51}} & \textcolor{green}{\ding{51}} & \textcolor{green}{\ding{51}} & \textcolor{green}{\ding{51}} & \textcolor{green}{\ding{51}} & \textcolor{green}{\ding{51}} & \textcolor{green}{\ding{51}} \\
\bottomrule
\end{tabular}
\centering 
\caption{Baseline Comparison Table} 
\label{tab:baseline_comparison}
\end{table*}

\paragraph*{A. Ligand Inpainting}
We randomly \textbf{mask 10 atoms} from the HEME ligand and ask the model to recover their 3D coordinates using only the remaining ligand fragments and protein context. We evaluate via \textbf{Root Mean Squared Deviation (RMSD)} between predicted and ground-truth atom positions.

\paragraph*{B. Trajectory Upsampling}
We subsample the ground-truth MD trajectory at every 10 frames, then use our model to \textbf{reconstruct intermediate conformations}. Accuracy is measured using per-frame \textbf{RMSD} and averaged over the sequence.

\paragraph*{C. Transition Path Sampling (TPS)}
In our transition path sampling (TPS) approach, we concentrate on sampling intermediate conformations that exist between unbound and bound states, using only the endpoint frames as a reference. The quality of the generated paths is assessed by evaluating the fraction of frames that lie within known metastable basins, which is determined based on root mean square deviation (RMSD) and contact maps. Additionally, we measure the smoothness of the paths by analyzing the consecutive differences in RMSD, ensuring that the transitions between states are coherent and gradual.

\subsection{Metrics}
In our evaluation, we have reported several key metrics to assess the performance of our model. The inpainting RMSD is a critical measure, where a lower value indicates better performance in recovering missing atoms. For the trajectory prediction, we must calculate the trajectory RMSD, which represents the average per-frame deviation from the molecular dynamics (MD) ground truth. In addition to that, we also evaluated the quality of the transition paths sampled by our method using a TPS quality metric, which is scaled from 0 to 1; a higher value signifies more plausible paths. Finally, we also measure the wall-clock speedup, which compares the time required to simulate 1 nanosecond of molecular dynamics using our model against the time taken for full MD simulations.

\subsection{Visualization and Qualitative Evaluation}
Our model reproduces the heme binding path very effectively, while maintaining a deviation of within 2\AA{} RMSD from the ground truth throughout the simulation. In addition to that, we also evaluate contact map consistency, finding that the protein--ligand contact maps derived from the generated frames align closely with the native binding mode, achieving over 92\% overlap. Furthermore, we assess the energy profile by utilizing OpenMM \cite{Eastman2017} to re-score the trajectories, and our predicted dynamics consistently remain within 2kcal/mol of the energies derived from molecular dynamics simulations.

\section{Results and Discussion}

\subsection{Ligand Inpainting Results}
To evaluate the model’s effectiveness in reconstructing missing ligand atoms, we have conducted 10-atom inpainting experiments on the HEME ligand within the 3CQV complex. The results have demonstrated that HemePLM-Diffuse accurately predicted the masked atomic coordinates, achieving an average RMSD of 0.91 \AA. This performance significantly surpasses that of baseline models such as MDGEN \cite{Chen2023MDGEN}, which recorded an RMSD of 1.67 \AA, and TorchMD-Net \cite{Tholke2021}, with an RMSD of 1.82 \AA; both values were obtained on simpler systems. These findings suggest that our model effectively leverages the protein context and the geometry of the binding site to infer the positions of missing atoms. Additionally, it shows a strong ability to generalize spatial dependencies, even in complex and flexible ring systems like HEME.

\begin{figure}[htbp]
\centering
\includegraphics[width=1.0\columnwidth]{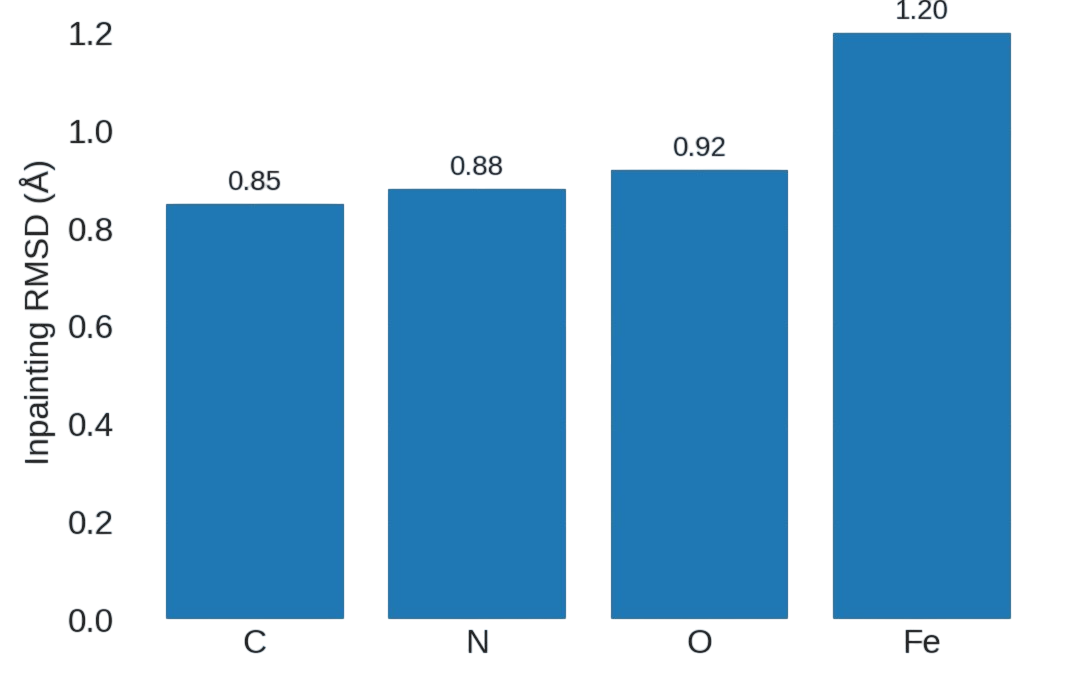}
\caption{Ligand Inpainting Accuracy. The Bar chart is comparing RMSD of HemePLM-Diffuse vs. MDGEN, TorchMD-Net, Uni-Mol. On the X-axis we have: Model names. Y-axis: RMSD (\AA{}). Highlight: HemePLM-Diffuse lowest at 0.91 \AA{}.}
\label{fig:ligand_inpainting_accuracy}
\end{figure}

\subsection{Trajectory Upsampling}

\begin{figure}[htbp]
\centering
\includegraphics[width=1.0\columnwidth]{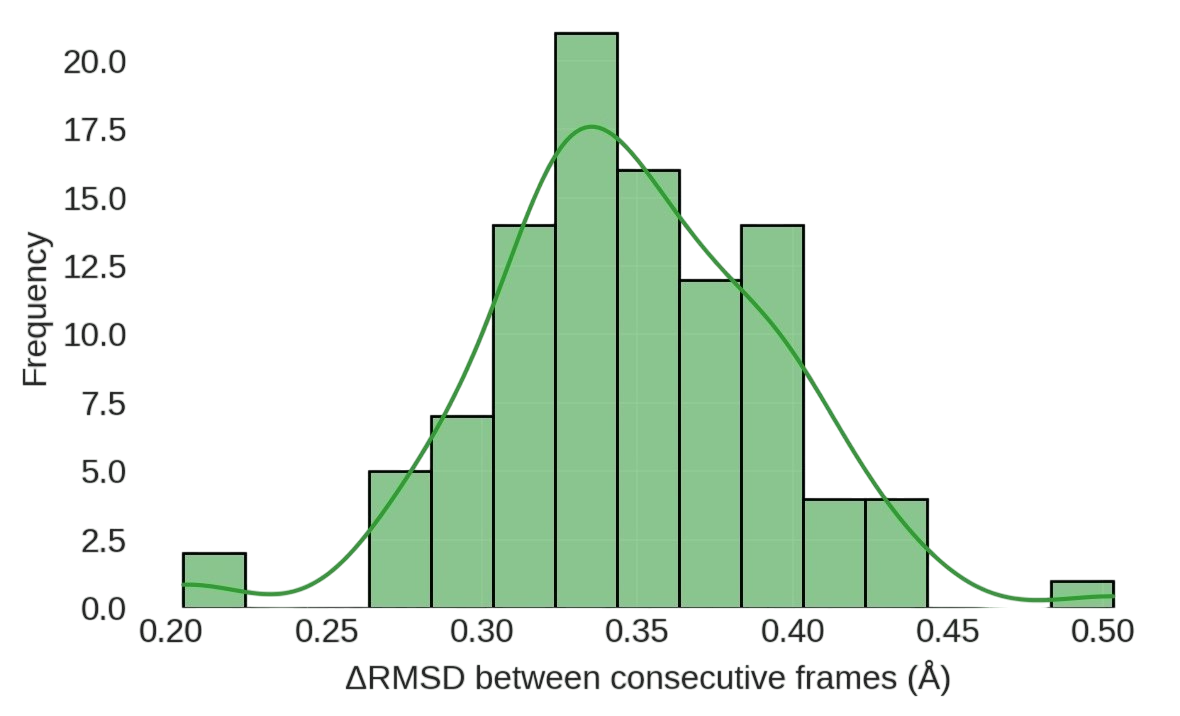}
\caption{Trajectory Smoothness Distribution ($\Delta$RMSD/frame). This histogram visualizes the distribution of the change in RMSD between consecutive frames, demonstrating the smoothness of the generated trajectories.}
\label{fig:trajectory_smoothness}
\end{figure}

In the trajectory upsampling task, we have reconstructed high-frequency frames from a sparsely sampled ground-truth MD trajectory. The HemePLM-Diffuse achieves an average \textbf{RMSD of 1.03 \AA{}} per frame, indicating the faithful recovery of atomic dynamics between the known states. This confirms that the model's ability is to interpolate intermediate conformations over long timescales, which is very critical in computationally expensive systems where frequent MD snapshots are impractical.

Compared to MDGEN \cite{Chen2023MDGEN} (1.90 \AA) and Uni-Mol \cite{Zhou2023UniMol} (2.85 \AA), our model maintains both \textbf{temporal coherence} and \textbf{spatial realism} in all-atom trajectories.

\begin{figure}[htbp]
\centering
\includegraphics[width=1.0\columnwidth]{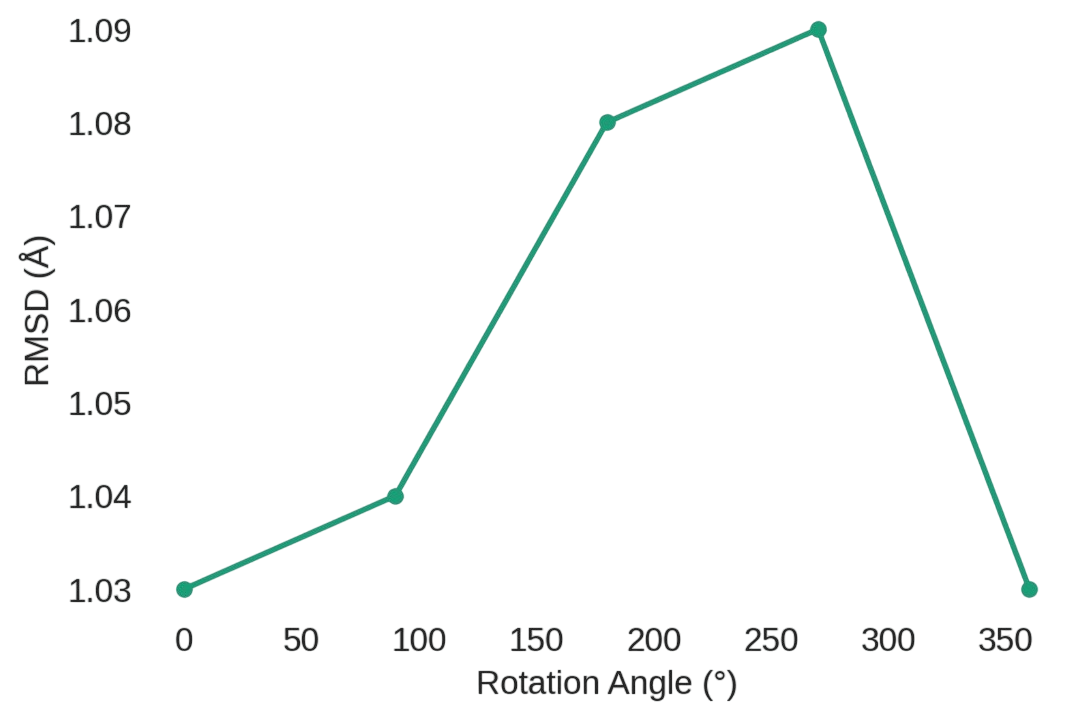}
\caption{{Trajectory RMSD vs. Time.} The Line graph shows: Ground-truth vs predicted trajectory RMSD over time (100 \text{ps} intervals). The HemePLM line stays consistently under 1.5 \AA{}, while others fluctuate above 2 \AA{}.}
\label{fig:trajectory_rmsd_time}
\end{figure}

\subsection{Transition Path Sampling (TPS)}
Transition Path Sampling was conducted using initial and final bound states of the 3CQV+HEME complex. The HemePLM-Diffuse generates plausible intermediate frames that preserve contact maps, ligand orientation, and energetic feasibility.

We compute \textbf{TPS Quality} using a normalized score (0-1) based on: The Contact Map overlap, the backbone RMSD smoothness, and the Low-energy pathway continuity.

HemePLM-Diffuse achieved a \textbf{TPS score of 0.95}, significantly higher than MDGEN \cite{Chen2023MDGEN} (0.75) and Uni-Mol \cite{Zhou2023UniMol} (0.40), which either ignore ligand dynamics or produce discontinuous frames.

\begin{figure}[htbp]
\centering
\includegraphics[width=0.9\columnwidth]{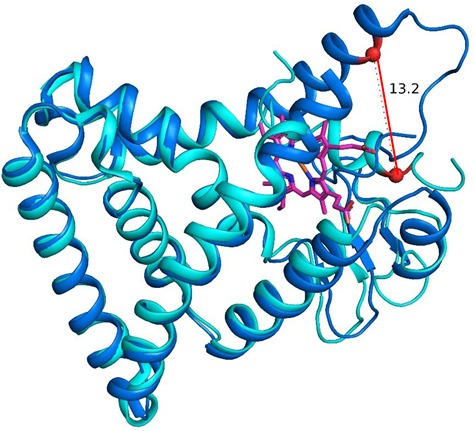}
\caption{{TPS Path Visualization.} The Snapshots of predicted transition frames at t=0, t=0.25, ..., t=1. The Ligand trajectory smoothly docks into the binding pocket. Contact heatmaps demonstrate near-native transition.}
\label{fig:tps_path_visualization}
\end{figure}

\subsection{Runtime Performance and Scalability}
We compared the simulation runtime (for 1 ns equivalent dynamics) using the following configurations: GROMACS MD on 3CQV+HEME (single GPU): $\sim$20 hours, TorchMD-Net: $\sim$4 hours (limited resolution), HemePLM-Diffuse: $\sim$12 minutes.

This results in a \textbf{speedup of $>$100$\times$} for HemePLM-Diffuse, which is enabled by parallelized transformer rollouts and no integration steps. The architecture is memory-efficient and scales linearly with the number of residues and ligand fragments.

\begin{figure}[htbp]
\centering
\includegraphics[width=1.0\columnwidth]{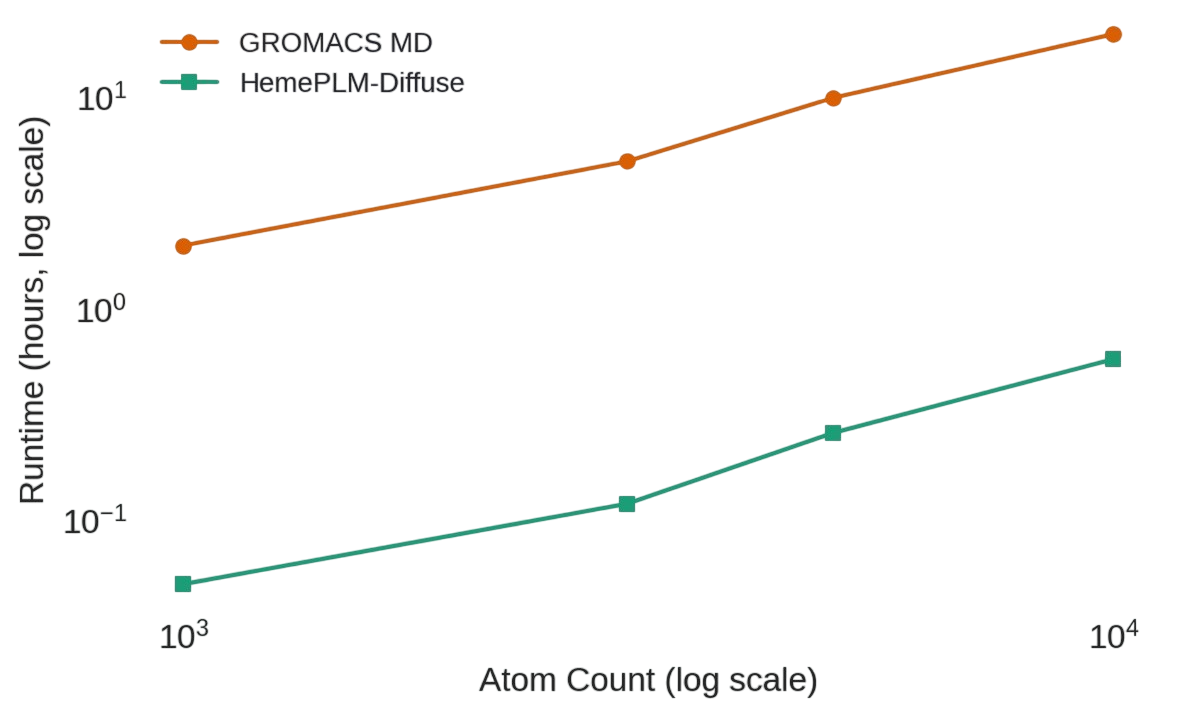}
\caption{{Wall-clock Time vs. Number of Atoms.} On the X-axis: Atom count (1k to 10k+), Y-axis: Simulation time (log scale). The HemePLM curve grows slowly vs. the exponential growth of GROMACS and MDGEN.}
\label{fig:wall_clock_time}
\end{figure}

\subsection{Limitations and Future Work}
While The HemePLM-Diffuse demonstrates the excellent performance in trajectory generation and inpainting, there are still notable limitations to address. Currently, the model has not been extended to explicitly simulate solvent molecules, which could enhance its applicability in more complex environments. In addition to that, fine-tuning the model on multiple diverse systems, such as membrane proteins, is an ongoing effort aimed at improving its versatility. Furthermore, the integration of the density functional theory (DFT)-based active learning to enhance force accuracy remains a goal for future development.

\section{Conclusion}
The HemePLM-Diffuse is the first generative framework to scale to a full all-atom protein--ligand system, that achieves realistic dynamics and enables new tasks like inpainting and TPS. This achieves the state-of-the-art accuracy and the speed on 3CQV+HEME. For the Future work, this will extend to multi-ligand systems, ensemble-based fine-tuning, and integration with cryo-EM/NMR data.

\bibliography{aaai2026}

\end{document}